\documentclass[11pt]{article}
\usepackage{moriond}
\usepackage{epsfig}

\def\Journal#1#2#3#4{{#1} {\bf #2}, #3 (#4)}

\def\NIMA{{\em Nucl. Instrum. Methods} A}

\def\PLB{{\em Phys. Lett.}  B}
\def\PRL{\em Phys. Rev. Lett.}
\def\PRD{{\em Phys. Rev.} D}

\def\be{\begin{equation}}
\def\ee{\end{equation}}
\def\bea{\begin{eqnarray}}
\def\eea{\end{eqnarray}}

\begin{document}
\vspace*{4cm}
\title{RESULTS FROM THE {\boldmath $\Upsilon$}(5S) ENGINEERING RUN (BELLE)}

\author{ A. DRUTSKOY }

\address{Physics Department, University of Cincinnati, 345 Clifton Court, \\
Cincinnati, OH 45221, USA}

\maketitle\abstracts{
We report results based on a 1.86\,fb$^{-1}$ data sample collected
on the $\Upsilon$(5S) resonance by the Belle 
detector at the KEKB asymmetric energy $e^+ e^-$ collider.
The inclusive production of $D_s$, $D^0$ and $J/\psi$ mesons 
at the $\Upsilon$(5S) is studied. From the $D_s$ inclusive 
branching fraction the ratio \mbox{$f_s = (16.4 \pm 1.4 \pm 4.1)\%$}
of $B_s^{(*)} \bar{B}_s^{(*)}$ to the total
$b\bar{b}$ quark pair production at the $\Upsilon$(5S) energy
is obtained in a model dependent way.
The exclusive decays $B_s \to D_s^{(*)+} \pi^- (/\rho^-)$
and $B_s \to J/\Psi \phi (/\eta)$ are studied and a significant 
$B_s$ signal is observed combining these modes. The $B_s$ meson 
production is found to proceed 
predominantly through the creation of $B^*_s \bar{B}^*_s$ pairs.
Upper limits on
$B_s \to K^+ K^-$, $B_s \to \phi \gamma$, $B_s \to \gamma \gamma$
and $B_s \to D_s^{(*)+} D_s^{(*)-}$ decays are reported.}

\section{Introduction}

The possibility of studying $B_s$ decays at very high luminosity
$e^+ e^-$ colliders running at the energy of the $\Upsilon$(5S) resonance
has been discussed in several theoretical papers \cite{teoa,teoc}.
In 2003 the CLEO experiment collected
0.42\,fb$^{-1}$ at the $\Upsilon$(5S) and observed
evidence for $B_s$ meson production in both 
inclusive \cite{cleoi} and exclusive \cite{cleoe} modes. 
However, simple calculations
assuming an approximate SU(3) symmetry indicate
that many interesting $B_s$ measurements require 
a data sample of at least $\sim$\,20\,fb$^{-1}$, which
is well within the capability of the $e^+ e^-$ $B$ factories.
To test the experimental feasibility of such measurements a data 
sample of 1.86\,fb$^{-1}$ was recently taken at the $\Upsilon$(5S) energy 
with the Belle detector \cite{belle} at the KEKB \cite{kekb}
asymmetric energy $e^+ e^-$ collider. 
This data sample is more than four times larger than the CLEO
dataset at the $\Upsilon$(5S).

An energy scan was performed just before the $\Upsilon$(5S) data taking
to define the position of the $\Upsilon$(5S) resonance production maximum.
An integrated luminosity of $\sim$\,30\,pb$^{-1}$ was collected
at five values of energy between 10825~MeV and 10905~MeV,
at intervals of 20~MeV. Finally, the data sample of $1.86\,\mathrm{fb}^{-1}$
was taken at the $\Upsilon$(5S) energy of $\sim$10869 MeV.
The experimental conditions of data taking at $\Upsilon$(5S)
are exactly the same as in $\Upsilon$(4S) or continuum runs.
The data sample of $3.67\,\mathrm{fb}^{-1}$ taken in the continuum
60 MeV in CM energy below the $\Upsilon$(4S) is used to estimate 
continuum backgrounds.



\section{Number of {\boldmath $b\bar{b}$} Events}

In the energy region of the $\Upsilon$(5S) the hadronic events
can be classified into three physics categories:
$u\bar{u}, d\bar{d}, s\bar{s}, c\bar{c}$ continuum events, 
$b\bar{b}$ continuum events and $\Upsilon$(5S)
events. The $b\bar{b}$ continuum and the $\Upsilon$(5S) 
events always produce final states with a pair of $B$ or 
$B_s$ mesons and, therefore, cannot be topologically separated.
We define the $b\bar{b}$ continuum and
$\Upsilon$(5S) events collectively
as $b\bar{b}$ events.
All $b\bar{b}$ events are expected to pass into the $B\bar{B}$,
$B\bar{B^*}$, $B^*\bar{B^*}$, $B\bar{B}\,\pi$, $B\bar{B}\,\pi \pi$,
$B\bar{B^*}\,\pi$, $B_s\bar{B_s}$, $B_s\bar{B_s^*}$ or 
$B_s^*\bar{B_s^*}$ final states.
The excited $B$ mesons decay to ground states via $B^* \to B \gamma$ and
$B_s^* \to B_s \gamma$ decay channels \cite{pdg}.

The $u\bar{u}, d\bar{d}, s\bar{s}, c\bar{c}$ continuum subtraction 
method is applied to obtain
the number of $b\bar{b}$ events in the $\Upsilon$(5S) data sample:

\vspace{-1.mm}
\begin{equation}
{\rm N}^{b\bar{b}}_{\rm 5S} = \frac{1}{\epsilon^{b\bar{b}}_{\rm 5S}} \left( N^{hadr}_{\rm 5S} - N^{hadr}_{\rm cont} \times \frac{{\cal L}_{\rm 5S}}{{\cal L}_{\rm cont}} \times \frac{E_{\rm cont}^{\,2}}{E_{\rm 5S}^{\,2}} \times \frac{\epsilon^{hadr}_{\rm 5S}}{\epsilon^{hadr}_{\rm cont}} \right)
\end{equation}

Here ${\rm N}^{b\bar{b}}_{\rm 5S}$ is the number of $b\bar{b}$
events in the $\Upsilon$(5S) data sample, and ${\rm N}^{hadr}_{\rm 5S}$
and $N^{hadr}_{\rm cont}$ are the numbers of hadronic events 
in the $\Upsilon$(5S) and continuum data samples, respectively.
The efficiency to select a $b\bar{b}$ event in the $\Upsilon$(5S) data 
sample, $\epsilon^{b\bar{b}}_{\rm 5S} = (99 \pm 1)\%$, and
the efficiency ratio for hadronic events in the $\Upsilon$(5S) 
and continuum data samples, 
$\epsilon^{hadr}_{\rm 5S} / \epsilon^{hadr}_{\rm cont} = 1.007 \pm 0.003$,
are obtained from MC simulation.
The integrated luminosity ratio ${\cal L}_{\rm 5S} / {\cal L}_{\rm cont}$ is
calculated using a standard Belle luminosity measurement procedure
based on the measurement of Bhabha events. 
The ratio ${\cal L}_{\rm 5S} / {\cal L}_{\rm cont} = 0.4740 \pm 0.0019$ 
is obtained after detailed calculations.
From Eq. (1) we obtain the number of $b\bar{b}$ events
in the $\Upsilon$(5S) data sample,
${\rm N}^{b\bar{b}}_{\rm 5S} = (5.61 \pm 0.03_{stat} \pm 0.29_{syst}) \times 10^5$.
The full systematic uncertainty of $\sim 5\%$ includes all 
systematic errors on parameters used
in Eq.\,(1) and is dominated by the uncertainty on the 
luminosity ratio definition.

\section{Inclusive {\boldmath $D_s$} Production}

The method of inclusive $D_s$ analysis of the $\Upsilon$(5S) data sample
was developed in \cite{cleoi}. Almost the same
technique is applied in this analysis.
The production of $D_s$ mesons at the $\Upsilon$(5S) is
studied in the clean decay mode, $D_s^+ \to \phi \pi^+$. 
The $D_s$ signals in the $\Upsilon$(5S) and continuum
data samples are seen in Fig.~1a for the normalized $D_s$ momentum
region $x(D_s) < 0.5$, where a $b\bar{b}$ contribution is expected.
The normalized momentum
is defined as $x(h) = P(h) / P_{max}(h)$, where $P(h)$ and $P_{max}(h)$ are
the measured momentum and the maximum possible momentum 
of particle $h$, respectively.
Continuum distributions are normalized to $\Upsilon$(5S) distributions
using the energy corrected luminosity ratio.
To extract the number of $D_s$ events, the candidate mass 
distribution is fitted
to a Gaussian to describe the signal and a linear function to describe 
the background.
The $x(D_s)$ distributions are shown in Fig.~1b
for the $\Upsilon$(5S) and continuum data samples.
These two distributions agree well in the region $x(D_s) > 0.5$, 
where $b\bar{b}$ events cannot contribute.
An excess of events in the region $x(D_s) < 0.5$ corresponds
to inclusive $D_s$ production in the $b\bar{b}$ events.

\begin{figure}[t!]
\vspace{-0.4cm}
\begin{center}
\epsfig{file=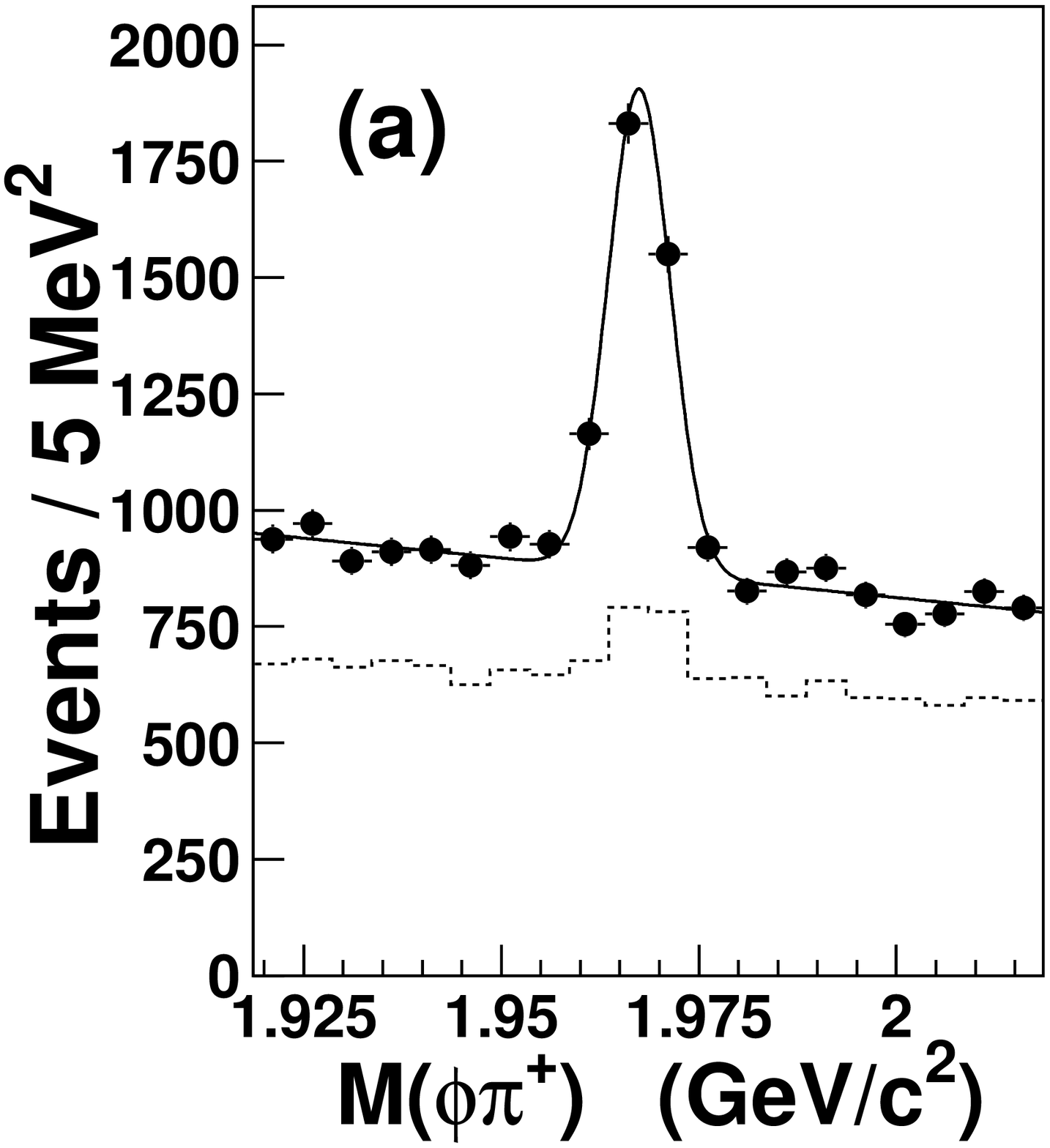,width=4.5cm,height=4.5cm}\epsfig{file=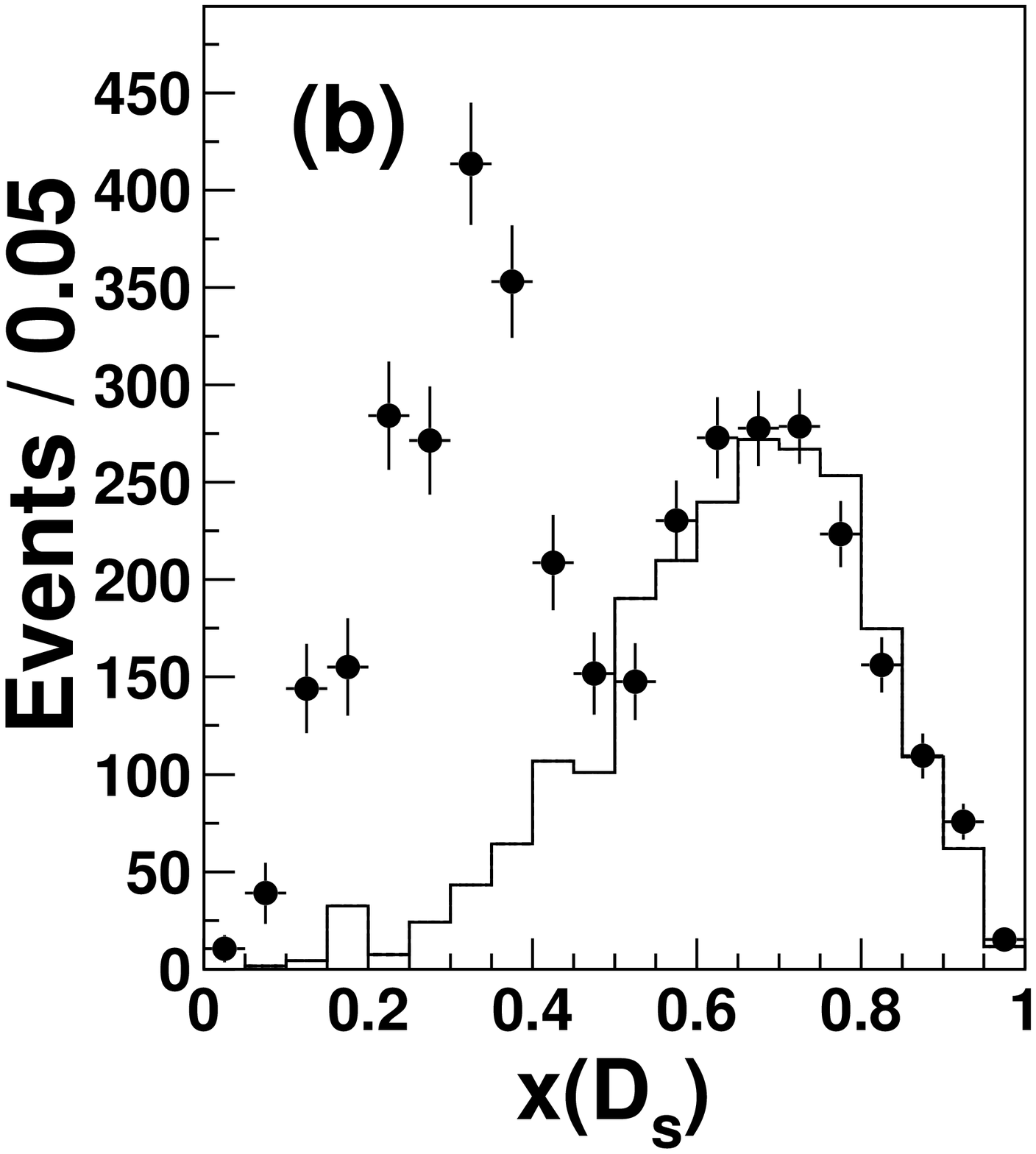,width=4.5cm,height=4.5cm}
\end{center}
\vspace{-0.3cm}
\caption{The $D_s$ signal (a) at the region $x(D_s) < 0.5$
for the $\Upsilon$(5S) (full circles with error bars) 
and continuum (hatched histogram) 
data samples, and the $D_s$ normalized momentum distributions (b) for the 
$\Upsilon$(5S) (full circles with error bars) and continuum (hatched 
histogram) data samples are shown.} 
\end{figure}

Subtracting the continuum contribution, applying
the efficiency correction, and summing over bins within the
interval $x(D_s) < 0.5$, the inclusive branching fraction
${\cal B}(\Upsilon$(5S)$\to D_s X) / 2 = (22.6 \pm 1.2 \pm 2.8)\%$
is obtained.
This value
is significantly larger than the branching fraction ${\cal B}(B \to D_s X)$,
which has been measured by the CLEO Collaboration \cite{cleoi}
to be $(9.0 \pm 0.3 \pm 1.4)\%$ and by the BaBar Collaboration 
\cite{babarb} to be 
$(8.94 \pm 0.16 \pm 1.12)\%$ (corrected for our value
of ${\cal B} (D_s^+ \to \phi \pi^+)$).
The significant increase of $D_s$ production at
$\Upsilon$(5S), compared with $\Upsilon$(4S), indicates a sizable 
$B_s$ production rate.

The ratio $f_s$ of 
$B_s^{(*)} \bar{B}_s^{(*)}$ events to all $b\bar{b}$ 
events at the $\Upsilon$(5S) can be obtained from equation:
\begin{equation}
{\cal B}(\Upsilon{\rm (5S)}\to D_s X) / 2 = f_s \cdot {\cal B}(B_s \to D_s X) +
(1 - f_s) \cdot {\cal B}(B \to D_s X)
\end{equation}
Using ${\cal B}(B_s \to D_s X) = (92 \pm 11)\%$ obtained 
in the reference \cite{cleoi} in a model dependent way
the value $f_s = (16.4 \pm 1.4 \pm 4.1)\%$ is obtained
from Eq. (2).
This value agrees well with
$f_s = (16.0 \pm 2.6 \pm 5.8)\%$
obtained by the CLEO collaboration
\cite{cleoi} from the $D_s$ inclusive analysis.
The dominant contributions to systematic uncertainty are the uncertainty on
${\cal B} (D_s^+ \to \phi \pi^+)$, the uncertainty 
on the number of $b\bar{b}$ events and the uncertainty on the model dependent 
assumption for ${\cal B}(B_s \to D_s X)$.

\section{Inclusive {\boldmath $D^0$} and {\boldmath $J/\psi$} Production}

The inclusive production of $D^0$ mesons at the $\Upsilon$(5S) is
studied in the decay mode $D^0 \to K^- \pi^+$ (Fig. 2a,b). 
Large $D^0$ signals are seen in Fig.~2a for the $\Upsilon$(5S)
and continuum data samples for the normalized momentum 
region $x(D^0) < 0.5$.
The number of $D^0$ mesons as a function of the normalized momentum $x(D^0)$
is shown in Fig.~2b for the $\Upsilon$(5S) and continuum data samples.
In a manner similar to the $D_s$ inclusive analysis,
the inclusive branching fraction 
${\cal B}(\Upsilon$(5S)$\to D^0 X) / 2 = (53.3 \pm 2.0 \pm 2.9)\%$ is 
determined.
Using the inclusive $D^0$ production branching fraction
of the $\Upsilon$(5S), $B$, and $B_s$ decays and replacing $D_s$ by
$D^0$ in Eq.(2),
the ratio $f_s = (18.7 \pm 3.6 \pm 6.7)\%$ of 
$B_s^{(*)} \bar{B}_s^{(*)}$ events to all $b\bar{b}$ events
at the $\Upsilon$(5S) is obtained.
This value agrees with our $D_s$ measurement.

\begin{figure}[b!]
\vspace{-0.4cm}
\begin{center}
\epsfig{file=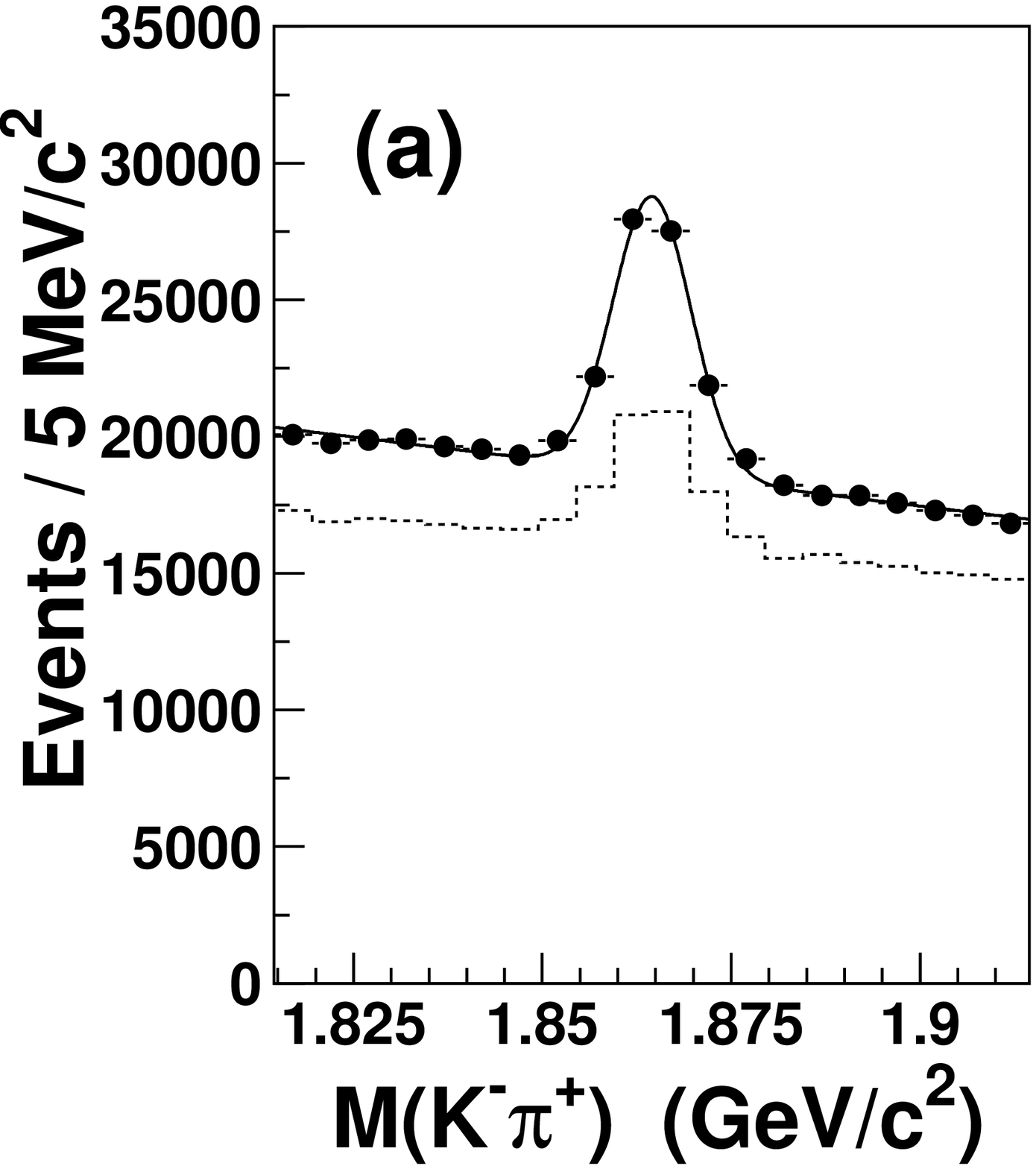,width=4.0cm,height=4.0cm}\epsfig{file=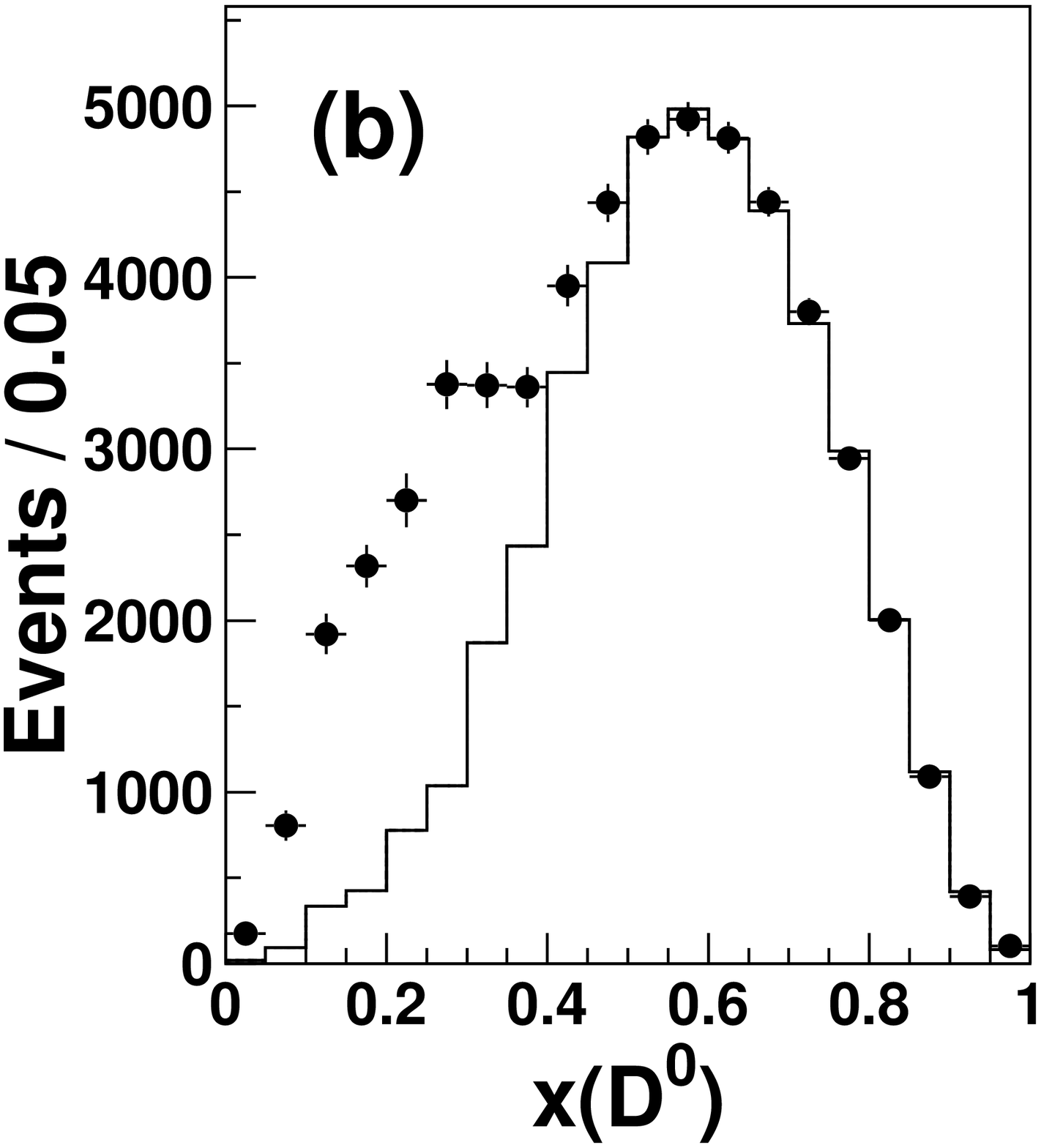,width=4.0cm,height=4.0cm}\epsfig{file=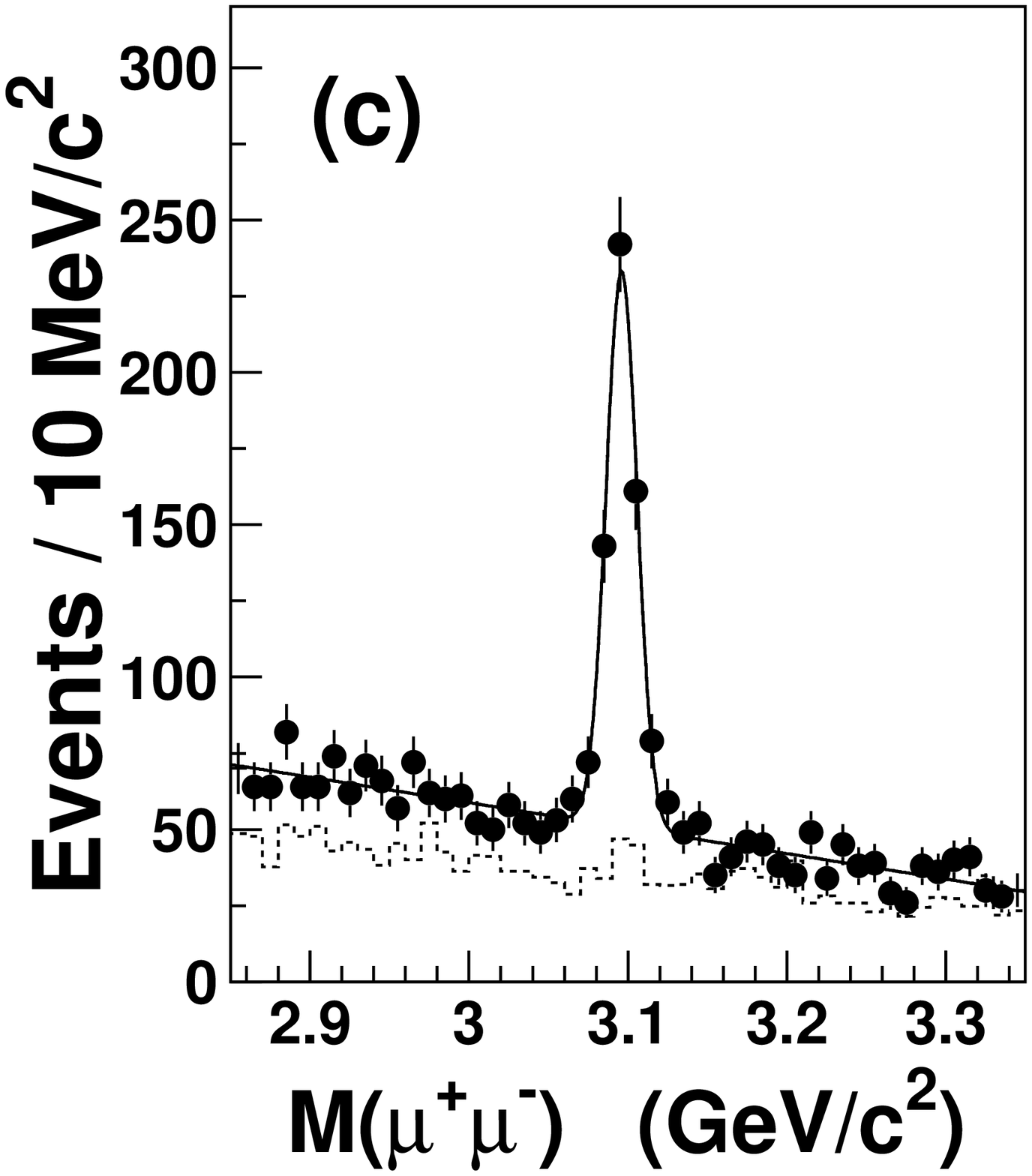,width=4.0cm,height=4.0cm}\epsfig{file=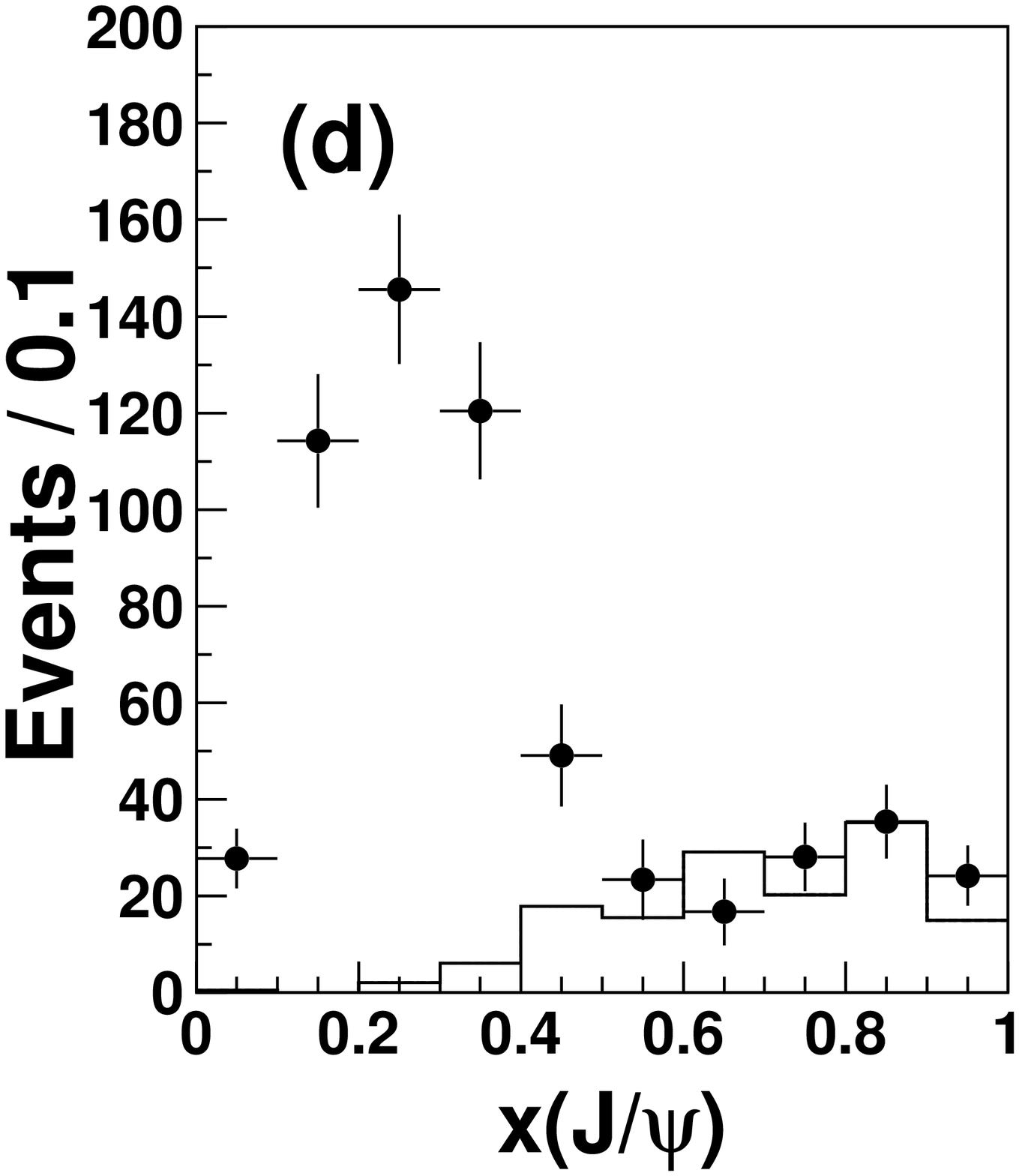,width=4.0cm,height=4.0cm}
\end{center}
\vspace{-0.3cm}
\caption{The $D^0$ (a) and $J/\psi$ (c) signals
at the region $x < 0.5$
for the $\Upsilon$(5S) (full circles with error bars) and continuum 
(hatched histogram) data samples, 
and the $D^0$ (b) and $J/\psi$ (d) normalized momentum 
distributions
for the $\Upsilon$(5S) (full circles with error bars) 
and continuum (hatched histogram) data samples are shown.}
\label{figall}
\end{figure}

The inclusive production of $J/\psi$ mesons is
studied in the decay mode $J/\psi \to \mu^+ \mu^-$ \mbox{(Fig.\,2c,d)}.
The $J/\psi$ production at the $\Upsilon$(5S) can be estimated
in a way similar to that used in the inclusive $D_s$ analysis.
In the region $x(J/\psi) < 0.5$ a prominent 
$J/\psi$ signal (Fig.~2c) is seen for the $\Upsilon$(5S) data sample,
whereas the $J/\psi$ signal in continuum is small.
The $J/\psi$ normalized momentum distributions are shown in Fig.~2d
for the $\Upsilon$(5S) and continuum data samples.

Using the PDG value \cite{pdg} 
${\cal B}(J/\psi \to \mu^+ \mu^-) = (5.88 \pm 0.10)\%$,
the inclusive branching fraction 
${\cal B}(\Upsilon$(5S)$\to J/\psi X) / 2 = (1.068 \pm 0.086 \pm 0.057)\%$ 
is obtained by summing events in the background subtracted and 
efficiency corrected $J/\psi$ normalized momentum distribution.
The $\Upsilon$(5S) branching fraction can be compared to the $B$ 
decay branching fraction \cite{pdg} 
${\cal B}(B \to J/\psi X) = (1.094 \pm 0.032)\%$,
because the inclusive $J/\psi$ production rate
in $B$ and $B_s$ decays is expected to be approximately equal.


\section{Exclusive {\boldmath $B_s \to D_s^{(*)+} \pi^-$},
{\boldmath $B_s \to D_s^{(*)+} \rho^-$} and 
{\boldmath $B_s \to J/\Psi \phi (/\eta)$} Decays}

In this analysis we studied the six conventional $B_s$ decay modes
$B_s \to D_s^+ \pi^-$, $B_s \to D_s^{*+} \pi^-$,
$B_s \to D_s^+ \rho^-$, $B_s \to D_s^{*+} \rho^-$,
$B_s \to J/\Psi \phi$ and $B_s \to J/\Psi \eta$,
which have large reconstruction efficiencies and
are described by unsuppressed conventional tree diagrams. 
$D_s^+$ mesons are reconstructed in the $\phi \pi^+$, $\bar{K}^{*0} K^+$
and $K_S^0 K^+$ decay channels.
To suppress continuum background we exploit topological cuts.


The signals can be observed using two variables: 
the energy difference $\Delta E\,=\,E^{CM}_{B_s}-E^{CM}_{\rm beam}$
and beam-constrained mass $M_{\rm bc}=
\sqrt{(E^{CM}_{\rm beam})^2\,-\,(p^{CM}_{B_s})^2}$;
$E^{CM}_{B_s}$ and $p^{CM}_{B_s}$ are the energy and momentum
of the $B_s$ candidate in the $e^+ e^-$ center-of-mass (CM) system
and $E^{CM}_{\rm beam}$ is the CM beam energy.
The $B_s$ mesons can be produced at the $\Upsilon$(5S) energy
via the intermediate $e^+ e^- \to B_s^{(*)} \bar{B}_s^{(*)}$ channels,
with $B_s^* \to B_s \gamma$.
Most $B_s$ signal events are concentrated within the ellipsoidal regions,
which are kinematically well separated for the $B_s^* \bar{B}_s^*$, 
$B_s^* \bar{B}_s$,
$B_s \bar{B}_s^*$ and $B_s \bar{B}_s$ channels. 

The distribution of data in $M_{\rm bc}$ and $\Delta E$
for the $B_s \to D_s^+ \pi^-$ decay mode is shown in Fig.~3a.
Nine events are observed within the $B_s$ signal ellipsoidal
region corresponding to $B_s^* \bar{B}_s^*$ pair production channel.
Background outside the signal regions is small.
The $M_{\rm bc}$ and $\Delta E$ scatter plots
are also obtained for the $B_s \to D_s^{*+} \pi^-$ (Fig.~3b),
$B_s \to D_s^{(*)+} \rho^-$ (Fig.~3c) and $B_s \to J/\Psi \phi /(\eta)$
(Fig.~3d) decays.
One of the observed $B_s \to J/\Psi \phi$ candidates
is reconstructed in the $J/\Psi \to \mu^+ \mu^-$ mode and one
in the $J/\Psi \to e^+ e^-$ mode.
One candidate is observed in the $J/\Psi \eta$ decay mode.

\begin{figure}[h!]
\vspace{-0.1cm}
\begin{center}
\epsfig{file=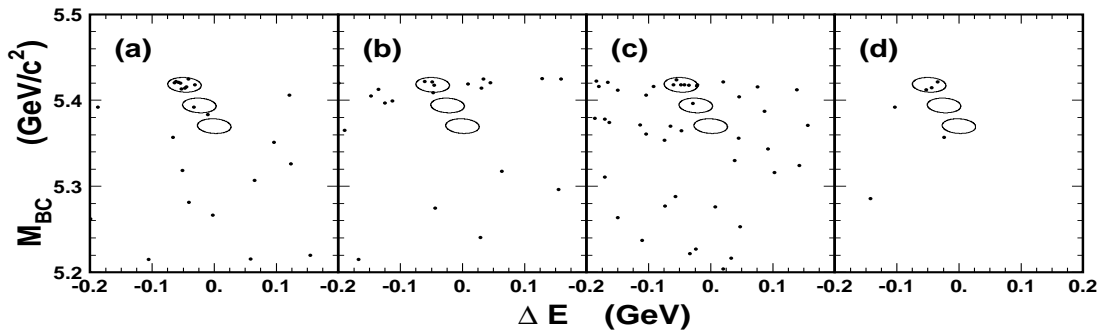,width=15.cm,height=5.cm}
\end{center}
\vspace{-0.3cm}
\caption{The $M_{\rm bc}$ and $\Delta E$ scatter plots
for the $B_s \to D_s^+ \pi^-$ (a),
$B_s \to D_s^{*+} \pi^-$ (b) 
and $B_s \to D_s^{(*)+} \rho^-$ (c) and $B_s \to J/\Psi \phi (\eta)$ (d)
decay modes are shown.}
\label{figall}
\end{figure}

The six $B_s$ modes shown in Fig.~3 are combined
to increase the statistical significance of the $B_s$
signal. Distributions in $\Delta E$ are obtained
separately for events from three $M_{\rm bc}$ intervals,
corresponding to $B_s$ production proceeding through the $B_s^* \bar{B}_s^*$, 
$B_s^* \bar{B}_s + B_s \bar{B}_s^*$ or $B_s \bar{B}_s$
channels, respectively.
Each of these three distributions is fitted by the sum of a Gaussian
to describe the signal and a linear function to describe background.
The fits yield $20.0 \pm 4.8$ events and
$1.3 \pm 2.0$ events for the $B_s^* \bar{B}_s^*$ and 
$B_s^* \bar{B}_s + B_s \bar{B}_s^*$ channels, respectively;
no events are observed in the $B_s \bar{B}_s$ channel.
From these numbers 
we obtain the ratio $\sigma (e^+ e^- \to B_s^* \bar{B}_s^*) /
\sigma (e^+ e^- \to B_s^{(*)} \bar{B}_s^{(*)}) = 0.94 ^{+0.06}_{-0.09}$
at the $\Upsilon$(5S) energy.
Potential models \cite{potb,potc} predict the 
fraction of $B_s^* \bar{B}_s^*$ channel over all these channels 
to be around 70$\%$.

The $B_s$ and $B_s^*$ masses can be extracted from the $M_{\rm bc}$ fits
in the $B_s^* \bar{B}_s^*$ channel, selecting 
candidates within the $-0.08 < \Delta E < -0.02\,$MeV range.
The $M_{\rm bc}$ distribution (Fig.~4a) is fitted by the sum of a Gaussian to
describe the signal and the so-called ARGUS function \cite{argus}
to describe background. The fit yields the mass value
M$(B_s^*) = (5418 \pm 1)\,$MeV/c$^2$. 
The observed width of the $B_s^*$ signal is $(3.6 \pm 0.6)\,$MeV/c$^2$ and 
agrees with the value obtained from the MC simulation, which assumes zero natural width.

\begin{figure}[h!]
\vspace{-0.1cm}
\begin{center}
\epsfig{file=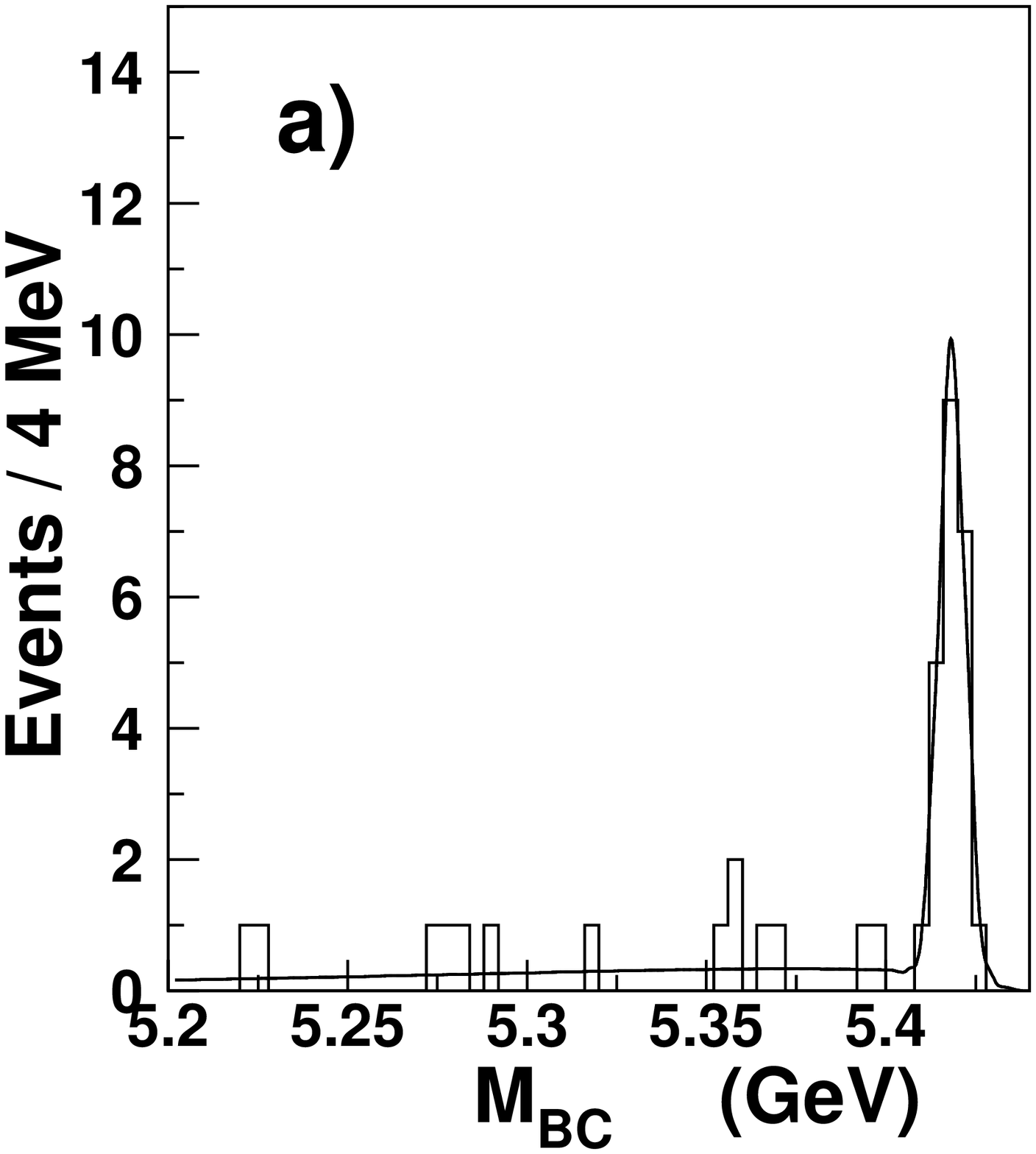,width=4.cm,height=4.cm}\epsfig{file=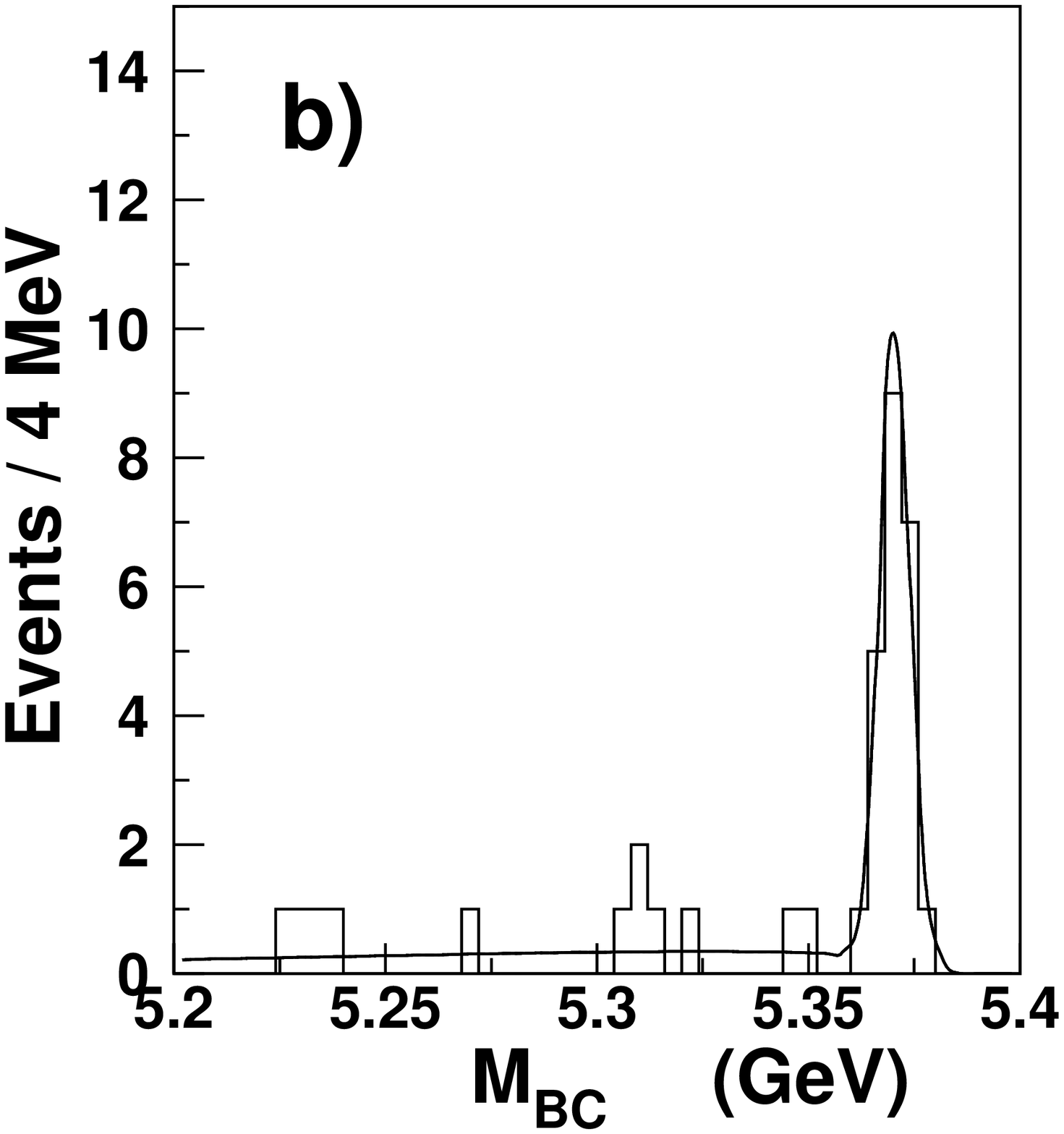,width=4.cm,height=4.cm}
\end{center}
\vspace{-0.3cm}
\caption{The $B_s^*$ (a) and $B_s$ (b) mass distributions for events
within the $-0.08 < \Delta E < -0.02\,$MeV interval, corresponding
to the $B_s^* \bar{B}_s^*$ channel.
Curves are obtained from the fits described in the text.}
\label{figall}
\end{figure}

Using events from the $B_s^* \bar{B}_s^*$ channel we can obtain
also the $B_s$ mass (Fig.~4b), replacing the 
energy $E^{CM}_{\rm beam}$ by
the term $E^{CM}_{\rm beam} - <\Delta E>$ in the mass formula.
The fit shown in Fig.~4b yields the $B_s$ mass
M$(B_s) = (5370 \pm 1 \pm 3)\,$MeV/c$^2$ and width
$\sigma(B_s) = (3.6 \pm 0.6)\,$MeV/c$^2$.
The second uncertainty in the $B_s$ mass value is the systematic
uncertainty due to the statistical uncertainty on the $<\Delta E>$ measurement.
The obtained $B_s$ mass agrees with
the most recent CDF measurement, M$(B_s) = (5366.0 \pm 0.8)\,$MeV/c$^2$.

\section{Rare {\boldmath $B_s$} Decays}

The distributions in data of $M_{\rm bc}$ and $\Delta E$ are obtained
for reconstructed $B_s \to \gamma \gamma$ (Fig.~5a), 
$B_s \to \phi \gamma$ (Fig.~5b), $B_s \to K^+ K^-$ (Fig.~5c) and 
$B_s \to D_s^{(*)+} D_s^{(*)-}$ (Fig.~5d) candidates.
Only the $B_s$ signal regions corresponding to the dominant
$B_s^* \bar{B}_s^*$ channel are indicated in Fig.~5.
These regions are wider for the $B_s \to \phi \gamma$ and
$B_s \to \gamma \gamma$ decays, where the energy losses
due to photon radiation lead to a large tail at lower values of 
$\Delta E$.

\begin{figure}[b!]
\vspace{-0.1cm}
\begin{center}
\epsfig{file=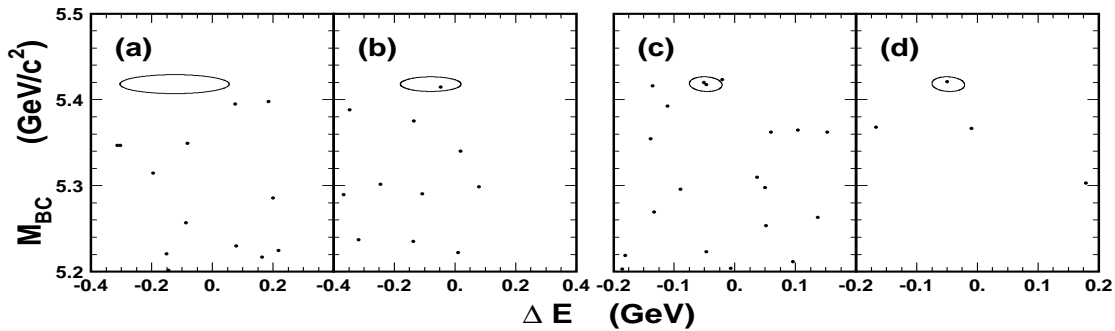,width=15.cm,height=5.cm}
\end{center}
\vspace{-0.3cm}
\caption{The scatter plots in $M_{\rm bc}$ and $\Delta E$
for the $B_s \to \gamma \gamma$ (a), $B_s \to \phi \gamma$ (b),
$B_s \to K^+ K^-$ (c) and
$B_s \to D_s^{(*)+} D_s^{(*)-}$ (d) decays.
The ellipses are indicating the $B_s$ signal 
regions for the $B_s^* \bar{B}_s^*$ channel.}
\label{figall}
\end{figure}

The numbers of events within the signal regions, estimated
background contributions, efficiencies, and upper limits
for these decays are listed in Table~1. For comparison,
the PDG upper limits are also shown. The number of 
initial $B_s^* \bar{B}_s^*$ pairs is obtained by 
multiplying the number of $B_s^{(*)} \bar{B}_s^{(*)}$ pairs
defined in the inclusive analysis by the production ratio
of $B_s^* \bar{B}_s^*$ pairs to $B_s^{(*)} \bar{B}_s^{(*)}$ pairs
obtained in analysis of conventional exclusive $B_s$ decays.
The efficiencies are determined from MC simulation.

\begin{table*}[h!]
\caption{The number of events in the signal region, estimated 
background contribution, efficiencies, upper limits
and PDG upper limits for the $B_s \to \gamma \gamma$, $B_s \to \phi \gamma$,
$B_s \to K^+ K^-$ and $B_s \to D_s^{(*)+} D_s^{(*)-}$ decay modes.}
\vspace{0.1cm}
\label{tab:bfr}
\begin{center}
\begin{tabular}
{|@{\hspace{0.3cm}}l@{\hspace{0.3cm}}|@{\hspace{0.3cm}}c@{\hspace{0.3cm}}|@{\hspace{0.3cm}}c@{\hspace{0.3cm}}|@{\hspace{0.3cm}}c@{\hspace{0.3cm}}|@{\hspace{0.3cm}}c@{\hspace{0.3cm}}|@{\hspace{0.3cm}}c@{\hspace{0.3cm}}|}
\hline
Decay mode & Yield, ev. & Backg., ev. & Eff. ($\%$) & {\it UL} ($10^{-4}$)& PDG {\it UL} ($10^{-4}$) \\
\hline
$B_s \to \gamma \gamma$ & 0 & 0.5 & 20.0 & 0.56 & 1.48 \\
$B_s \to \phi \gamma$ & 1 & 0.15 & 5.9  & 4.1 & 1.2 \\
$B_s \to K^+ K^-$ & 2 & 0.14 & 9.5 & 3.4 & 0.59 \\
$B_s \to D_s^+ D_s^-$ & 0 & 0.02 & 0.020 & 710. & - \\
$B_s \to D_s^{*+} D_s^-$ & 1 & 0.01 & 0.0099 & 1270. & - \\
$B_s \to D_s^{*+} D_s^{*-}$ & 0 & $<$0.01 & 0.0052 & 2730. & - \\
\hline
\end{tabular}
\end{center}
\vspace{-0.3cm}
\end{table*}

The obtained upper limit for the $B_s \to \gamma \gamma$ decay
is about three times smaller, than the current best world value.
Within the Standard Model this decay is expected to proceed via an intrinsic 
penguin diagram \cite{gga} and the branching fraction
is expected to be \mbox{$(0.5-1.0) \times 10^{-6}$}.
However that branching fraction is sensitive to some Beyond
Standard Model (BSM) contributions and can be higher by one to two orders
of magnitude in some BSM models \cite{ggc,ggd}.
Important results can be obtained using
the $B_s \to D_s^{(*)+} D_s^{(*)-}$ decay modes \cite{dsds}. These
modes are expected to be predominantly $CP$ eigenstates
and, because of expected large branching fractions, should
lead to a sizable lifetime difference between $CP$-odd and $CP$-even $B_s$ mesons.
Assuming the sum of these branching fractions
to be around (5-8)$\%$, we can expect about 5 events in each
of four possible final states with statistics of $\sim$30\,fb$^{-1}$.
Within the SM framework such measurement can provide 
an important constraint on the value of $\Delta \Gamma_s / \Gamma_s$.

\section*{Acknowledgments}
We gratefully
acknowledge NSF award number PHY-0611671 for travel support.

\end{document}